\documentclass[aps,superscriptaddress,10pt,pra,a4paper,twocolumn]{revtex4-1}

\usepackage[english]{babel}
\usepackage{verbatim}
\usepackage{natbib}

\newcommand{\bra}[1]{\langle#1|}
\newcommand{\ket}[1]{|#1\rangle}

\newcommand{\eket}[1]{|\tilde{#1}\rangle}
\newcommand{\stsp}{STP}

\def\2#1{{\cal #1}}

\usepackage{graphicx}
\usepackage{amssymb,amsmath,amsfonts,amsthm}
\usepackage{booktabs}

\begin{document} 

\title{Quantum Transport Enhancement by Time-Reversal Symmetry Breaking} 

\author{Zolt\'an Zimbor\'as}
\affiliation{Institute for Scientific Interchange, Via Alassio 11/c, 10126 Torino, Italy}
\affiliation{Department of Theoretical Physics, University of the Basque Country 
  UPV/EHU, P.O. Box 644, E-48080 Bilbao, Spain}
\author{Mauro Faccin}
\affiliation{Institute for Scientific Interchange, Via Alassio 11/c, 10126 Torino, Italy}
\author{Zolt\'an K\'ad\'ar}
\affiliation{Institute for Scientific Interchange, Via Alassio 11/c, 10126 Torino, Italy}
\author{James Whitfield}
\affiliation{Institute for Scientific Interchange, Via Alassio 11/c, 10126 Torino, Italy}
\affiliation{Vienna Center For Quantum Science and Technology, Boltzmanngasse 5 
  1090 Vienna, Austria}
\author{Ben Lanyon}
\affiliation{Institut f\"ur Quantenoptik und Quanteninformation, 
  Otto-Hittmair-Platz 21a 6020 Innsbruck, Austria}
\author{Jacob Biamonte}
\email{jacob.biamonte@qubit.org}
\affiliation{Institute for Scientific Interchange, Via Alassio 11/c, 10126 Torino, Italy}
\affiliation{Centre for Quantum Technologies, National University of Singapore, 
  Block S15, 3 Science Drive 2, Singapore 117543}

\begin{abstract}
Quantum mechanics still provides new unexpected effects when considering the transport of energy and information. Models of continuous time quantum walks, which implicitly use time-reversal symmetric Hamiltonians, have been intensely used to investigate the effectiveness of transport. 
Here we show how breaking time-reversal symmetry 
of the unitary dynamics in
this model can enable directional control, enhancement, and suppression of quantum transport. Examples ranging from exciton transport to complex networks are presented. 
This opens new prospects
for more efficient methods to transport energy and information. \\
\end{abstract}

\maketitle 

Understanding quantum transport is key to developing more robust communication networks, more effective energy transmission, and improved information processing devices. Continuous time quantum walks have become a
standard model to study and understand quantum transport phenomena
\cite{FG98,CCDFGS03,MB11,Kempe03,Kendon06,Salvador12,BB12}. 
Time-reversal symmetric (TRS) Hamiltonians have
characterized all quantum walk models to date.
This symmetry
implies that the site-to-site transfer probability at time $t=T$ is the same as at time $t=-T$, 
thereby prohibiting directional biasing.
Here we introduce and study continuous time ``chiral'' quantum walks whose
dynamics break TRS\@. Our findings show that the breaking of TRS offers the
possibility of directional biasing in the unitary dynamics and allows one to suppress or enhance
transport relative to the standard quantum walk. 
One subtlety of this effect is that time-reversal asymmetry cannot affect the site-to-site transport 
in some simple cases, such as linear chains and trees---this is proven in the 
Methods Section. Prior efforts in the area of quantum transport have focused on controlling and 
directing  transport using either \textit{in situ}
tunable Hamiltonians \cite{Godsil, Burgarth,Xiang} or tailoring specific initial 
states \cite{Eisfeld}. In contrast to known approaches, we consider states 
initially prepared in the standard \textit{site-basis} and time-independent 
Hamiltonians that induce time-asymmetric evolutions in the unitary part of their dynamics.

While the effect of TRS breaking dynamics in the context of
quantum walks has not been investigated, it has been studied intensely 
in the condensed matter literature. These investigations range 
from the very early work of Peierls \cite{Peierls}, through the famous examples of the
Hofstadter butterfly \cite{HB} and the Quantum Hall \cite{QHall}  effect,
 up to recent research on 
TRS breaking in topological insulators \cite{HK10} and on artificial gauge fields in 
optical lattice potentials \cite{DGJO11}. In contrast to the present study, these works always concentrated on  
many-body dynamics in regular lattices, while
in the context of quantum walks, one is instead interested in characteristically different 
scenarios: e.g.~the dynamics of
single individual particles or excitons (usually starting from a single site) 
moving on complicated networks (sometimes with a bath included). The examples we 
study are from a variety of modern research topics (e.g.\ 
photosynthetic exciton transport and complex networks) and
considerably extend the domain of application of known results about TRS breaking beyond
solid state applications.

To demonstrate the effect of TRS breaking, we chose five
examples which illustrate the main ideas of directionality, suppression and
enhancement of transport. 
The first example is a unitary quantum switch where the phase, that is, the time 
reversal asymmetry parameter controls the direction of quantum transport. 
The second example examines
transport in a linear chain of triangles, showing a 633\% transport
speed-up for the chiral quantum walk.
In connection with this, we also demonstrate complete suppression of
chiral quantum walks on loops with an even number of sites.
We then consider a system widely studied in the exciton transport literature: 
the Fenna-Matthew-Olsen complex (FMO). 
Although this naturally occurring system is highly efficient, we find that
the introduction of chiral terms allows for an enhancement of transport speed by
$7.68$\%.
It has recently been shown that the effect does appear in similar light harvesting complexes \cite{Engel}.
{Finally, to investigate the robustness of the effect of TRS breaking on 
transport, we consider randomly generated small-world networks.}
By appending time-reversal asymmetric terms to only the edges of the network
connected to the final site,
we could increase the speed of the site-to-site
transport on these randomly generated graphs significantly, up to 130\%.

\section*{Results}

In the standard literature on continuous time quantum walks \cite{FG98,CCDFGS03, MB11,Kempe03, Kendon06}, the time-independent walk Hamiltonian is defined by a real weighted adjacency matrix $J$ of an underlying undirected graph, 
\begin{equation}
H_{QW} = \sum^{sites}_{n,m} J_{nm}(\ket{n}\bra{m} +\ket{m}\bra{n}) \, .
\end{equation} 
The condition that the hopping weights $J_{nm}$ are real numbers 
implies that the induced transitions between two sites
are symmetric under time inversion. We can break this symmetry while maintaining the hermitian property of the operator by appending a complex phase to an edge: $J_{nm}\rightarrow J_{nm}e^{i\theta_{nm}}$ resulting in a continuous time \textit{chiral quantum
walk} (CQW) governed by 
\begin{equation}\label{eqn:cqw} 
  H_{CQW} = \sum_{n,m} J_{nm}e^{i \theta_{nm}} \ket{n}\bra{m} + 
    J_{nm}e^{-i \theta_{nm}}\ket{m}\bra{n} 
    \,.
\end{equation} 
When acting on the single exciton subspace the Hamiltonian given in 
Eq. \eqref{eqn:cqw} can be expressed in terms of the spin-half Pauli matrices:
\begin{align}
  H_{CQW}=&
  \sum_{n,m} J_{nm}\cos(\theta_{nm})(\sigma^x_{n}\sigma^x_{m} 
  +\sigma^y_{n}\sigma^y_{m}) \notag\\
  &
  +\sum_{n,m} J_{nm}\sin(\theta_{nm})(\sigma_{n}^x\sigma^y_{m} -\sigma^y_{n}\sigma^x_{m})
\end{align}
which arises in a variety of physical systems when magnetic fields are considered. We explore 
a proof-of-concept experimental demonstration of this effect in Supplementary Information, Section S2.

In the CQW framework, we investigate coherent quantum dynamics and incoherent dynamics within the Markov approximation. Both types of evolution are
included in the Lindblad equation \cite{Kossakowski72,Lindblad76,Breuer02,Whitfield10}:
\begin{align}
	\frac{d}{dt}\rho(t)=&
        \mathcal{L}\{\rho\} \notag
        =
        -i[H_{CQW}:\rho]\\
        &+\sum_k L_k \rho L_k^\dag-\frac{1}{2}\left(L_k^\dag L_k\rho+\rho L_k^\dag L_k\right)
\end{align}
where $\rho(t)$ is the density operator describing the state of the system at time $t$ and $L_k$ are Lindblad operators inducing stochastic jumps between quantum states. For example, using the usual terminology of Markovian processes, we call site $t$ a trap if it is coupled to site $s$ by the Lindblad jump operators, $L_k=\ket{t}\bra{s}$. 
The site-to-site transfer probability, $P_{n\rightarrow m}(t)=\bra{m}\rho(t)\ket{m}$, gives the occupancy probability of site $m$ at time $t$ with initial condition $\rho(0)=\ket{n}\bra{n}$. 
Note that the present study, while utilizing open system dynamics, is not  related to the enhancement of transport due to 
quantum noise \cite{SMPE12,MRLA08} which has been well studied in the context of photosynthesis \cite{MRLA08,lloyd2011}. Here the emphasis is instead on the effect the breaking time-reversal symmetry of the Hamiltonian dynamics can have on transport.

To quantify the transport properties of quantum walks, we use the 
\textit{half-arrival time}, $\tau_{1/2}$, as the earliest time when the occupancy probability of the target site is one half.  
We will also
make use of the transport speed, $\nu_{1/2}$, defined as  the reciprocal of $\tau_{1/2}$.
\begin{figure}
  \centering
  \includegraphics[width=0.8\columnwidth]{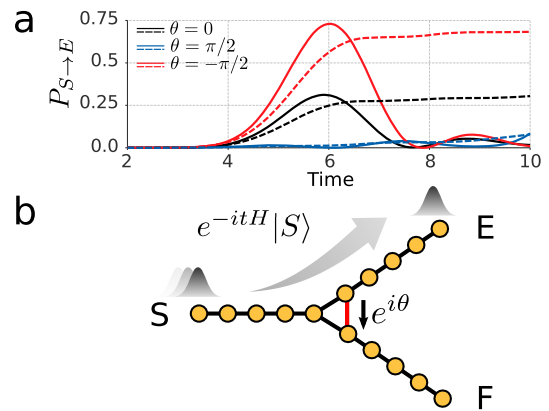}
  \caption{The quantum switch. ({\bf a}) Directional biasing: enhanced transport in the preferred
  direction. ({\bf b}) The plot shows the occupancy probability $P_{S\to E}$
of site $E$ with the particle initially starting from site $S$ with and without sink 
(dashed and solid lines, respectively). This evolution is time-reversal 
asymmetric as replacing $t$
  with $-t$ results in the particle moving from site $S$ towards site $F$.
  When starting at site $E$, the particle evolves towards site
  $F$.  
  By replacing  $t$ with $-t$,  a particle initially at site $E$
  evolves towards the initial configuration ({\bf b}).  To recover time-reversal
  symmetric  transition probabilities in the evolution ({\bf b}), 
requires that one also performs the antiunitary operation
  \cite{W31} on the Hamiltonian mapping $\theta$ to $-\theta$.  This has the same effect as
  reflecting the configuration horizontally across the page while leaving the site labels intact.}
  \label{fig:switch}
\end{figure}

We now introduce a quantum switch which enables directed
transport and could, in principle, be used to create a logic gate and offer
future implementations of transport devices to store and process energy and
information. Fig.~\ref{fig:switch} presents an example of this switch.
The value of a phase ($e^{i \theta }$)  appended to a single control edge across the 
junction allows selective biasing of transport through the switch. The maximal 
biasing occurs at $|\theta|=\pi/2$, and the sign determines the direction.
The first maxima of $P_{S\rightarrow E}(t)$ (transfer probability from site S to E) 
in the unitary dynamics without traps  can be enhanced by 134\% or suppressed to 91\% 
with respect to the non-chiral case. When considering traps in the Lindbladian evolution, 
the optimal transport efficiency is 81.4\% in the preferred direction. 
The switch violates TRS as $P_{S\to E}(-t)\neq P_{S\to E}(t)$. By using $P_{S\to E}(-t)=P_{E\to S}(t)$ and the symmetry of the configuration 
$P_{E\to S}(t)=P_{S\to F}(t)$, we conclude that transport is biased towards the
opposite pole when running backwards in time, see Fig~\ref{fig:switch}.
Note that the behaviour of the switch is largely independent of the length of the connecting wires.

\begin{figure}
  \begin{center}
    \includegraphics[width=\columnwidth]{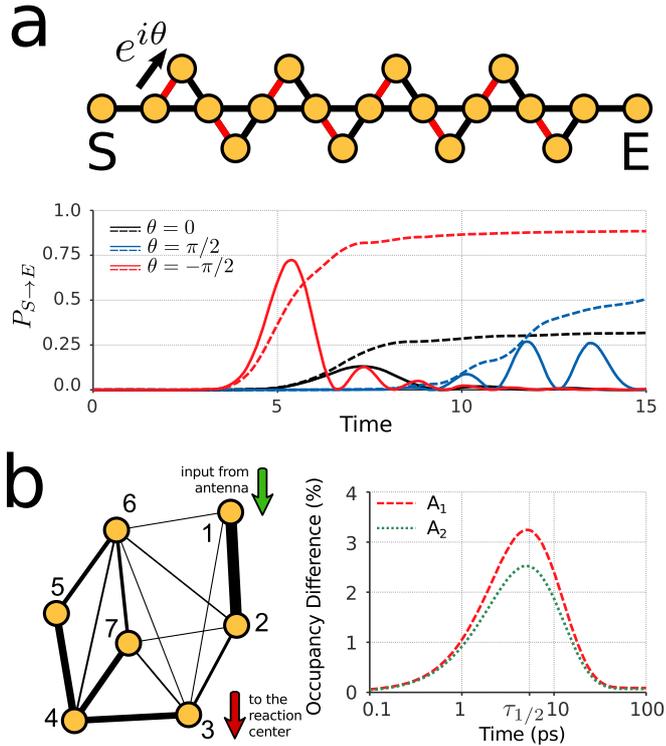}
  \end{center}
  \caption{
    ({\bf a}) Triangle chain and ({\bf b}) the FMO complex.
    ({\bf a}) The phase $e^{i \theta}$ is applied to the red edges simultaneously in the triangle chain. 
    The plot illustrates the occupancy probability at the end site $E$ as a 
    function of time for different values 
    of the phase $\theta$ with and without trapping (dashed and solid lines, respectively). 
    ({\bf b}) shows the occupancy difference with respect to the time reversal symmetric
    Hamiltonian of the FMO complex. We use an optimization procedure to enhance the transport.
    While holding the magnitude of the couplings constant, we optimize two sets of phases, 
    $A_1$ and $A_2$, which correspond to seven and three edges with an enhancement at
    $\tau_{1/2}$ of $3.25$\% and $2.25$\%, respectively.
  }
  \label{fig:saw+fmo}
\end{figure}

We will now utilize the directional biasing of the triangle to give an example of a
speed-up of chiral walks.
Using the composition of eight triangular switches as depicted in
Fig.~\ref{fig:saw+fmo}a, by simultaneously varying all phases along the red
control edges to the same value, we examine the effect of time-reversal
asymmetry on transport.
We find that the occupation probability as a function of $\theta$ is symmetric
about $\pm \pi/2$ with the negative value corresponding to maximal
enhancement and the positive value to maximal suppression.
Unlike the occupation probability maxima in the switch, here the first apexes
are separated in time.
When we include trapping, the half-arrival time is reduced from
the non-chiral value $\tau_{1/2}=38.1$ to $5.2$ which represents a $633$\% enhancement.
 To conclude this section we focus on suppression of transport by chiral quantum walks. 
A good example is the polygon with an even number of sites.
In this case complete suppression can be achieved by appending a phase of $\pi$
to one of the links in the cycle; thereby rendering it impossible for the quantum
walker to move to the diametrically opposite site. 
This is a discrete space version 
of a known effect in Aharonov-Bohm loops \cite{Datta}.
The proof that the site-to-site transfer probability is zero in this case for all times
also in our discrete-space and open-system walks can be found in the 
Methods Section. However, note that the discrete 
even-odd effect, which implies that only loops comprised of odd particles can exhibit transport enhancement, and only even loops may exhibit complete suppression, has no known continuous analog.

In natural and synthetic excitonic networks such as photosynthetic complexes
and solar cells, we are faced with non-unitary quantum evolution due to
dissipative and decoherent interaction with the environment. Studies have
shown that dissipative quantum evolution surpasses both classical and purely
quantum transport (for interesting recent examples see \cite{Whitfield10,SMPE12}).
A widely studied process of such dissipative exciton transport is the
one occurring in the Fenna-Matthews-Olsen complex (FMO), which connects the
photosynthetic antenna to a reaction centre in green sulphur
bacteria \cite{MRLA08,caruso09,fleming10,ringsmuth2012}. Due to the low light exposure of these bacteria, there is evolutionary pressure to optimize exciton transport. 
Therefore, the site energies and site-to-site couplings in the
 system are evolutionarily optimized, yielding a highly efficient transport \cite{lloyd2011}.
However, it is an open question whether or not there occurs 
time-reversal asymmetric hoping terms in these systems, and whether these are
optimized. Recent 2D Electronic Spectroscopy results lead to the conclusion that , e.g., 
in the light harvesting complex LH2 hopping terms with
complex phases are indeed present \cite{Engel}.
Here we ask whether such TRS breaking interactions
 may further enhance the efficiency of the light harvesting process.
We consider the traditional real-hopping Hamiltonian modeling
transport on the FMO, and allow for TRS breaking by introducing complex phases and find that 
the transport speed can be further increased.
We study the seven site model of the FMO using an open system description
that includes the thermal bath, trapping at the reaction centre, and
recombination of the exciton \cite{MRLA08,caruso09, plenio08}. 
By performing a standard optimization procedure (as outlined in the
Supplementary Information, Section  S3) that
varies the phase on a subset of seven edges, we found a combination of phases
where the transport speed, $\nu_{1/2}$, is enhanced by $7.68$\%. In
Fig.~\ref{fig:saw+fmo}b, the enhancement of the time dependent occupation probability is shown for the chiral quantum walk. 
We note that optimization over only three edges already changes the transport speed 
by 5.92\%, see Supplementary Information, Section  S3.

Complex network theory has been used in abstract studies of quantum information science; see for example \cite{Acin, Acin2}.  Here we turn to the theory of complex networks to determine if optimization procedures limited
to small subsets of edges will generally lead to improved transport in larger
and possibly randomly generated networks. 
We found a positive answer when testing 
the site-to-site transport between oppositely aligned nodes in the
Watts-Strogatz model~\cite{WS98}. 
This family of small-world networks continuously connects a class of 
regular cyclic graphs to that of completely random networks
(Erd\H{o}s-R\'enyi models\cite{ER60}) by changing the value of the rewiring probability. 

\begin{figure}
  \centering
  \includegraphics[width=\columnwidth]{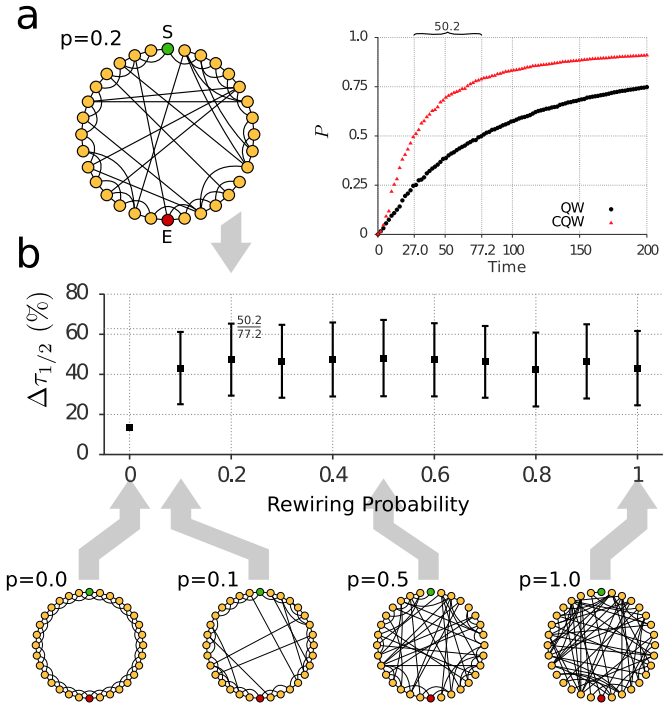}
  \caption{Transport enhancement of the chiral quantum walk is robust 
    across randomly generated Watts-Strogatz networks.
    An example of this small-world network, with rewiring probability $p=0.2$, is
    depicted in ({\bf a}).
    The transfer probability $P$ from site $S$ to the sink
    connected to site $E$ is plotted in a realization of the network.
    ({\bf b}) shows the average enhancement of half arrival 
    time ($\Delta\tau_{1/2}$) for different values of $p$.
  }\label{fig:WS}
\end{figure}

We numerically investigated graphs with 32 nodes, average degree four and range
over rewiring probability $p$ considering 200 different graph realizations for each value of $p$.
An example with $p=0.2$ is depicted in Fig.~\ref{fig:WS}a.
Here the occupancy of a sink connected to site $E$ is compared between the chiral
walk and its achiral counterpart.
The particle begins at site $S$ and we perform the optimization of the phases
only on edges connected to site $E$.
In the case of the chiral quantum walk, the sink reaches half-occupancy in 54.8\% less time on average.  

\section*{Discussion}

In all the examples studied,  we found that 
the effect that TRS breaking  has on transport is 
non-trivially affected by the topology of the network. 
In this regard, a key observation is the following.
If two Hamiltonians are related by on-site unitary transformations mapping
$\ket{n}$ to $e^{i\alpha_n}\ket{n} $, then the site-to-site transition 
probabilities will be identical.
This fact provides a tool to reduce the effective space of
phase parameters for controlling transport. In the Methods Section, we provide a more formal treatment of this symmetry of 
the site-to-site transition probabilities.
For instance, we prove that the site-to-site transfer probability is 
insensitive to phases in tree graphs. 
For bipartite graphs the phases can have an effect, however,
the dynamics still remains time-reversal symmetric.

A further consequence is that the sums of phases along a chosen orientation 
of a loop are the unique invariants under on-site unitary transformations.
For example, placing
phases on the edges of the triangle loop of the quantum switch is equivalent to placing the sum of them
on just one edge.
In a wide range of cases and particularly in all examples we considered, we found strong evidence of the robustness the effect has on transport. 
For instance, the examples in Fig.~\ref{fig:WS} shows that in the Watts-Strogatz model, the transport
enhancement due to the time reversal asymmetry of the Hamiltonian is insensitive
to changes of the rewiring probability $p$ and the clustering coefficient measuring the density of triangles in the graph.
Finally, additional calculations show that scale free networks such as the Barab\'asi-Albert model \cite{BA99},
show a similar transport enhancement, indicating robustness also with respect
to the degree distribution.

This study pioneers the exploration of a new degree of freedom
that allows for a significant improvement of control in the
engineering of quantum transport. The fact that we were able to optimize and 
control transport by adjusting the
phase on only a few edges inside a complex network and that the effect was 
relevant in a host of examples adds optimism to the
robustness of this approach.
Experimental demonstrations of 
the effects we predict are within reach of existing
hardware, as outlined in the Supplementary Information, Section  S2. \\

\section*{Methods}

\subsection{Analytical methods}

\paragraph{Site-to-site transfer probability}
 The Markovian open-system dynamics of a continuous time chiral quantum walk is 
given by the Kossakowski-Lindblad equation \cite{Kossakowski72, Lindblad76,
Breuer02,Whitfield10}
\begin{align}
  \mathcal{L}\{\rho\}=&
  -i[H_{CQW},\rho] \notag\\
  &+
    \sum_{(n,m)} c_{mn}\left(L_{nm}\rho L_{nm}^\dag
    -\frac 12 \{L_{nm}^\dag L_{nm},\rho\}
    \right)\, ,
\label{LE}
\end{align}
where the chiral Hamiltonian $H_{CQW}$ is defined in Eq.~(2),
and the Lindblad operators are given as  $L_{mn}=\ket{m}\bra{n}$ with $c_{nm}\geq 0$.
Transport from vertex $\ket{S}$ to vertex $\ket{E}$ during such dynamics is
characterized by the site-to-site transfer probability (\stsp). In the 
unitary case ($c_{nm}=0$)  it is given by
\begin{equation} \label{ArrT}
P_{S\to E}(t)=\mbox{Tr}\,(e^{-iH_{CQW}t} \rho_S\,e^{iH_{CQW}t}
\rho_E)
\end{equation}
with $\rho_S=\ket{S}\bra{S}$ and
$\rho_E=\ket{E}\bra{E}$, while for the general Markovian case
it is
\begin{equation}
  P_{S\rightarrow E}(t)=
  \mbox{Tr}(e^{\mathcal{L}t}\{\rho_S\}\rho_E)\ . \label{tevL}
\end{equation}

\paragraph{Time-reversal symmetry of the unitary achiral dynamics}
In the setting of quantum walks, the time-reversal operator $T$ acts 
as complex conjugation (with respect to the vertex basis) \cite{W31}:
\begin{equation*}
T \sum_{v \in V} \alpha_v | v \rangle = 
\sum_{v \in V} \alpha^*_v | v \rangle\ .
\end{equation*}
The antiunitarity of $T$ and  $T^2={\mathbf 1}$ implies that 
$T^\dagger=T$.
The time-reversal of a Hamiltonian $H$ is given as $THT^\dagger(=THT)$. The
$H\mapsto THT$ action
is represented in parameter space  by the replacement
$\theta_{mn}\mapsto-\theta_{mn}$ in Eq.~(2). Thus exactly the achiral quantum walks
are left invariant by this action.
The \stsp 's  of $H$ ($P_{S\to E}(t)$) and that of $H'=THT$ ($P'_{S\to E}(t)$)
are related in the following way:
\begin{equation*}
  P'_{S\to E}(t)= P_{S\to E}(-t)\;\mbox{and}\;P'_{S\to E}(t)=P_{E\to S}(t),
\end{equation*}
which can be verified using $T\rho_vT=\rho_v$ and the
cyclicity of the trace as follows:
\begin{align*}
  P'_{S\to E}(t) &= 
    \mbox{Tr} (e^{-i(THT)t}\rho_S\, e^{i(THT)t}\rho_E)\\ 
    &=\mbox{Tr}(Te^{iHt} T\rho_S\, T e^{-iHt} T\rho_E)\\
    &=\mbox{Tr} (e^{iHt}T\rho_S\, T e^{-iHt} T\rho_E T)\\
    &=\mbox{Tr}  (e^{iHt} \rho_S\,   e^{-iHt} \rho_E)=
    P_{S\to E}(-t) \, ,\\
  P_{S\to E}(-t)&=
    \mbox{Tr} (e^{iHt} \rho_S\, e^{-iHt} \rho_E)\\
    &=\mbox{Tr} (e^{-iHt} \rho_E\, e^{iHt} \rho_S)=
    P_{E\to S}(t)\ .
\end{align*}  
A crucial consequence of the above is that in the case of achiral 
quantum walks, the transition probabilities are the same at time $t$ and $-t$,
i.e. $P_{S \to E}(t)=P_{S \to E}(-t)$, and directional biasing is 
prohibited $P_{S\to E}(t) = P_{E\to S}(t)$. However, $H\neq THT^\dag$ does
not necessarily imply that transition rates are asymmetric in time.
This is because $THT^\dag$ might be gauge-equivalent to $H$, as will be seen in
the next section.

\paragraph{Gauge transformations}
Formal gauge transformations, already introduced in
the early work of Peierls\cite{Peierls}, are useful 
tools to study our models. Such a transformation is simply
a local change of basis, i.e., a diagonal unitary
\begin{equation}
  U_d \ket{n} = e^{i\alpha_{n}} \ket{n}\ \label{gt}.
\end{equation} 
Here we collect a few of its properties and generalize them for the case of
open systems with a Markovian bath. For us the starting point will be that
it leaves the \stsp\ invariant. 
To prove this, let us first note that any unitary basis-change $U$ would
induce a transformation on the Lindblad superoperator $\mathcal{L} \to \mathcal{L}'$ with
\begin{equation*} 
\mathcal{L}'\{\rho \} = U\mathcal{L}\{U^{\dagger} \rho U\}U^{\dagger} \, .
\end{equation*}
Using this and  the invariance of localized states under diagonal unitaries
($U_d^{\dagger} \rho_{v} U_d=\rho_{v}$), we arrive at
\begin{align*} 
  P'_{S\to E}(t) &= 
  \mbox{Tr} (e^{\mathcal{L'}t}\{\rho_S\}\rho_E)\\
  &=\mbox{Tr}(U_d e^{\mathcal{L}t}\{U_d^{\dag}\rho_S U_d\}U_d^\dagger\rho_E)\\
  &=\mbox{Tr}(e^{\mathcal{L}t}\{\rho_S\}U_d^{\dag}\rho_EU_d)\\
  &=\mbox{Tr}(e^{\mathcal{L}t}\{\rho_S\}\rho_E)=P_{S\to E}(t) \, ,
\end{align*}
which proves the invariance of the \stsp\ under the gauge transformations
defined by Eq.~\eqref{gt}.

\begin{figure}
    \centering
    \includegraphics[width=\columnwidth]{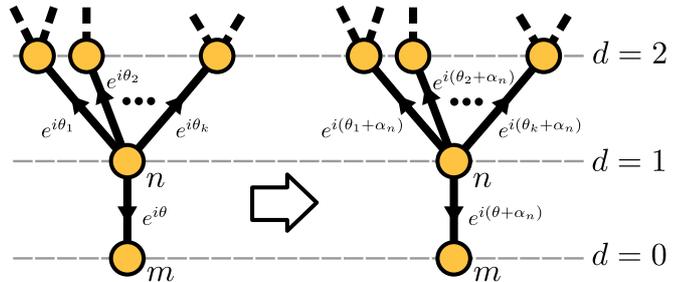}
    \caption{Effect of the gauge transformation $\ket{n}\to
    e^{i\alpha_n}\ket{n}$ on vertex $n$.
    Phases on edges can be gauge-transformed without changing the transition
    amplitudes, as described in the text.
    Here we arrange the graph as a tree rooted at $d=0$.
    }
  \label{fig:gauge-tree}
\end{figure}

Under these diagonal transformations,
the parameters of the quantum walk Hamiltonian transform as 
\begin{equation}
\theta_{mn}\mapsto \theta_{mn}+\alpha_m-\alpha_n\ ,\label{gtl} 
\end{equation}
as illustrated in Fig. \ref{fig:gauge-tree}.
The incoherent part of the Lindblad equation \eqref{LE} does not change
since the Lindblad operators transform as $L_{nm} \to e^{i(\alpha_m-\alpha_n)}L_{nm}$ and these
phases cancel in Eq.~\eqref{LE}, since $L_{nm}$ and $L^\dag_{nm}$ appear
paired. 
Two important properties of the model now follow: 
($i$) phases on tree graphs can be transformed out completely and 
($ii$) the sum of phases along loops is invariant under gauge transformations. 

The first property is illustrated in Fig.~\ref{fig:gauge-tree}. 
Let us take an arbitrary tree graph and pick a vertex $m$ with only one
neighbour.
Redraw every other vertex on successive levels characterized by the distance
$d$ of the vertexes from the given vertex $m$.
Note that the number of edges connecting two vertices, $d$, is by definition,
unique in tree graphs.
In such an arrangement only one edge emanates downwards from a given vertex on a line
of $d>0$ so Fig.~\ref{fig:gauge-tree} represents the general neighbourhood of a vertex $n$
having distance $d=1$ from $m$. 
The indicated gauge transformation with 
$\alpha_n=-\theta$ removes the phase from the bottom edge. 
Then, one iterates the procedure for all vertices at level $d=2$ and
consecutively for all levels. In this way, all phases are removed.  
For the second property, pick an orientation on a loop of $N$
vertexes  and compute $\theta:=\sum_{i=1}^N\phi_{i,i+1}$,
considering $\phi_{N,N+1}\equiv\phi_{N,1}$. A gauge 
transformation  $\ket{n}\mapsto e^{i\alpha_n}\ket{n}$, according to
Eq.~\eqref{gtl} leads to:
\begin{equation*}
  \phi_{n,n+1}\mapsto\phi_{n,n+1}+\alpha_n,\quad \phi_{n-1,n}\mapsto \phi_{n-1,n}-\alpha_n,
\end{equation*}
so the sum $\theta$ remains unaffected. 

\subsection{Numerical methods}

We used the Quantum Information Toolkit \cite{Ville}, a 
software package for the Matlab programming language.
The optimization procedure used in the FMO and the Watts-Strogatz examples rely on the
{\it Interior-point Optimization} algorithm of the Matlab minimization tool kit.
We start the optimization procedure several times from different randomly
chosen points of the parameter space, to reach the global minimum of
cost function with more certainty.
Source code for all simulations done in this work is available upon request.


\bibliographystyle{apsrev}

\begin{thebibliography}{42}
\expandafter\ifx\csname natexlab\endcsname\relax\def\natexlab#1{#1}\fi
\expandafter\ifx\csname bibnamefont\endcsname\relax
  \def\bibnamefont#1{#1}\fi
\expandafter\ifx\csname bibfnamefont\endcsname\relax
  \def\bibfnamefont#1{#1}\fi
\expandafter\ifx\csname citenamefont\endcsname\relax
  \def\citenamefont#1{#1}\fi
\expandafter\ifx\csname url\endcsname\relax
  \def\url#1{\texttt{#1}}\fi
\expandafter\ifx\csname urlprefix\endcsname\relax\def\urlprefix{URL }\fi
\providecommand{\bibinfo}[2]{#2}
\providecommand{\eprint}[2][]{\url{#2}}

\bibitem[{\citenamefont{Farhi and Gutmann}(1998)}]{FG98}
\bibinfo{author}{\bibfnamefont{E.}~\bibnamefont{Farhi}} \bibnamefont{and}
  \bibinfo{author}{\bibfnamefont{S.}~\bibnamefont{Gutmann}},
  \bibinfo{journal}{Phys. Rev. A} \textbf{\bibinfo{volume}{58}},
  \bibinfo{pages}{915} (\bibinfo{year}{1998}).

\bibitem[{\citenamefont{{Childs} et~al.}(2003)\citenamefont{{Childs}, {Cleve},
  {Deotto}, {Farhi}, {Gutmann}, and {Spielman}}}]{CCDFGS03}
\bibinfo{author}{\bibfnamefont{A.~M.} \bibnamefont{{Childs}}},
  \bibinfo{author}{\bibfnamefont{R.}~\bibnamefont{{Cleve}}},
  \bibinfo{author}{\bibfnamefont{E.}~\bibnamefont{{Deotto}}},
  \bibinfo{author}{\bibfnamefont{E.}~\bibnamefont{{Farhi}}},
  \bibinfo{author}{\bibfnamefont{S.}~\bibnamefont{{Gutmann}}},
  \bibnamefont{and} \bibinfo{author}{\bibfnamefont{D.~A.}
  \bibnamefont{{Spielman}}}, \bibinfo{journal}{Proc. 35th Annual ACM STOC. ACM,
  NY} pp. \bibinfo{pages}{59--68} (\bibinfo{year}{2003}).

\bibitem[{\citenamefont{{M{\"u}lken} and {Blumen}}(2011)}]{MB11}
\bibinfo{author}{\bibfnamefont{O.}~\bibnamefont{{M{\"u}lken}}}
  \bibnamefont{and} \bibinfo{author}{\bibfnamefont{A.}~\bibnamefont{{Blumen}}},
  \bibinfo{journal}{Phys. Rep.} \textbf{\bibinfo{volume}{502}},
  \bibinfo{pages}{37} (\bibinfo{year}{2011}).

\bibitem[{\citenamefont{{Kempe}}(2003)}]{Kempe03}
\bibinfo{author}{\bibfnamefont{J.}~\bibnamefont{{Kempe}}},
  \bibinfo{journal}{Contemp. Phys.} \textbf{\bibinfo{volume}{44}},
  \bibinfo{pages}{307} (\bibinfo{year}{2003}).

\bibitem[{\citenamefont{{Kendon}}(2006)}]{Kendon06}
\bibinfo{author}{\bibfnamefont{V.}~\bibnamefont{{Kendon}}},
  \bibinfo{journal}{Math. Struct. in Comp. Sci} \textbf{\bibinfo{volume}{17}},
  \bibinfo{pages}{1169} (\bibinfo{year}{2006}).

\bibitem[{\citenamefont{Venegas-Andraca}(2012)}]{Salvador12}
\bibinfo{author}{\bibfnamefont{S.}~\bibnamefont{Venegas-Andraca}},
  \bibinfo{journal}{Quantum Information Processing} pp. \bibinfo{pages}{1--92}
  (\bibinfo{year}{2012}), ISSN \bibinfo{issn}{1570-0755}.

\bibitem[{\citenamefont{Baez and Biamonte}(2012)}]{BB12}
\bibinfo{author}{\bibfnamefont{J.~C.} \bibnamefont{Baez}} \bibnamefont{and}
  \bibinfo{author}{\bibfnamefont{J.}~\bibnamefont{Biamonte}},
  \emph{\bibinfo{title}{A course on quantum techniques for stochastic
  mechanics}} (\bibinfo{year}{2012}), \bibinfo{note}{235 pages},
  \eprint{1209.3632}.

\bibitem[{\citenamefont{Godsil and Severini}(2010)}]{Godsil}
\bibinfo{author}{\bibfnamefont{C.}~\bibnamefont{Godsil}} \bibnamefont{and}
  \bibinfo{author}{\bibfnamefont{S.}~\bibnamefont{Severini}},
  \bibinfo{journal}{Phys. Rev. A} \textbf{\bibinfo{volume}{81}},
  \bibinfo{pages}{052316} (\bibinfo{year}{2010}).

\bibitem[{\citenamefont{Burgarth et~al.}(2011)\citenamefont{Burgarth,
  D'Alessandro, Hogben, Severini, and Young}}]{Burgarth}
\bibinfo{author}{\bibfnamefont{D.}~\bibnamefont{Burgarth}},
  \bibinfo{author}{\bibfnamefont{D.}~\bibnamefont{D'Alessandro}},
  \bibinfo{author}{\bibfnamefont{L.}~\bibnamefont{Hogben}},
  \bibinfo{author}{\bibfnamefont{S.}~\bibnamefont{Severini}}, \bibnamefont{and}
  \bibinfo{author}{\bibfnamefont{M.}~\bibnamefont{Young}},
  \bibinfo{journal}{arXiv:} \textbf{\bibinfo{volume}{1111.1475v1}}
  (\bibinfo{year}{2011}).

\bibitem[{\citenamefont{Xiang et~al.}(2013)\citenamefont{Xiang, Litinskaya,
  Shapiro, and Krems}}]{Xiang}
\bibinfo{author}{\bibfnamefont{P.}~\bibnamefont{Xiang}},
  \bibinfo{author}{\bibfnamefont{M.}~\bibnamefont{Litinskaya}},
  \bibinfo{author}{\bibfnamefont{E.~A.} \bibnamefont{Shapiro}},
  \bibnamefont{and} \bibinfo{author}{\bibfnamefont{R.~V.} \bibnamefont{Krems}},
  \bibinfo{journal}{New Journal of Physics} \textbf{\bibinfo{volume}{15}},
  \bibinfo{pages}{063015} (\bibinfo{year}{2013}).

\bibitem[{\citenamefont{Eisfeld}(2011)}]{Eisfeld}
\bibinfo{author}{\bibfnamefont{A.}~\bibnamefont{Eisfeld}}, \bibinfo{journal}{J.
  Chem. Phys.} \textbf{\bibinfo{volume}{379}}, \bibinfo{pages}{33}
  (\bibinfo{year}{2011}).

\bibitem[{Pei(1993)}]{Peierls}
\bibinfo{journal}{Z. Phys.} \textbf{\bibinfo{volume}{80}}, \bibinfo{pages}{763}
  (\bibinfo{year}{1993}).

\bibitem[{\citenamefont{Hofstadter}(1976)}]{HB}
\bibinfo{author}{\bibfnamefont{D.~R.} \bibnamefont{Hofstadter}},
  \bibinfo{journal}{Physical review B} \textbf{\bibinfo{volume}{14}},
  \bibinfo{pages}{2239} (\bibinfo{year}{1976}).

\bibitem[{\citenamefont{Sarma and Pinczuk}(2008)}]{QHall}
\bibinfo{author}{\bibfnamefont{S.~D.} \bibnamefont{Sarma}} \bibnamefont{and}
  \bibinfo{author}{\bibfnamefont{A.}~\bibnamefont{Pinczuk}},
  \emph{\bibinfo{title}{Perspectives in quantum Hall effects}}
  (\bibinfo{publisher}{John Wiley \& Sons}, \bibinfo{year}{2008}).

\bibitem[{\citenamefont{Hasan and Kane}(2010)}]{HK10}
\bibinfo{author}{\bibfnamefont{M.~Z.} \bibnamefont{Hasan}} \bibnamefont{and}
  \bibinfo{author}{\bibfnamefont{C.~L.} \bibnamefont{Kane}},
  \bibinfo{journal}{Reviews of Modern Physics} \textbf{\bibinfo{volume}{82}},
  \bibinfo{pages}{3045} (\bibinfo{year}{2010}).

\bibitem[{\citenamefont{Dalibard et~al.}(2011)\citenamefont{Dalibard, Gerbier,
  Juzeli{\=u}nas, and {\"O}hberg}}]{DGJO11}
\bibinfo{author}{\bibfnamefont{J.}~\bibnamefont{Dalibard}},
  \bibinfo{author}{\bibfnamefont{F.}~\bibnamefont{Gerbier}},
  \bibinfo{author}{\bibfnamefont{G.}~\bibnamefont{Juzeli{\=u}nas}},
  \bibnamefont{and}
  \bibinfo{author}{\bibfnamefont{P.}~\bibnamefont{{\"O}hberg}},
  \bibinfo{journal}{Reviews of Modern Physics} \textbf{\bibinfo{volume}{83}},
  \bibinfo{pages}{1523} (\bibinfo{year}{2011}).

\bibitem[{\citenamefont{Harel and Engel}(2012)}]{Engel}
\bibinfo{author}{\bibfnamefont{E.}~\bibnamefont{Harel}} \bibnamefont{and}
  \bibinfo{author}{\bibfnamefont{G.~S.} \bibnamefont{Engel}},
  \bibinfo{journal}{Proceedings of the National Academy of Sciences}
  \textbf{\bibinfo{volume}{109}}, \bibinfo{pages}{706} (\bibinfo{year}{2012}).

\bibitem[{\citenamefont{Kossakowski}(1972)}]{Kossakowski72}
\bibinfo{author}{\bibfnamefont{A.}~\bibnamefont{Kossakowski}},
  \bibinfo{journal}{Rep. Math. Phys} \textbf{\bibinfo{volume}{3}},
  \bibinfo{pages}{247} (\bibinfo{year}{1972}).

\bibitem[{\citenamefont{Lindblad}(1975)}]{Lindblad76}
\bibinfo{author}{\bibfnamefont{G.}~\bibnamefont{Lindblad}},
  \bibinfo{journal}{Commun. Math. Phys} \textbf{\bibinfo{volume}{48}},
  \bibinfo{pages}{119} (\bibinfo{year}{1975}).

\bibitem[{\citenamefont{Breuer and Petruccione}(2002)}]{Breuer02}
\bibinfo{author}{\bibfnamefont{H.-P.} \bibnamefont{Breuer}} \bibnamefont{and}
  \bibinfo{author}{\bibfnamefont{F.}~\bibnamefont{Petruccione}},
  \emph{\bibinfo{title}{The theory of open quantum systems}}
  (\bibinfo{publisher}{Oxford University Press}, \bibinfo{year}{2002}).

\bibitem[{\citenamefont{Whitfield et~al.}(2010)\citenamefont{Whitfield,
  Rodriguez-Rosario, and Aspuru-Guzik}}]{Whitfield10}
\bibinfo{author}{\bibfnamefont{J.~D.} \bibnamefont{Whitfield}},
  \bibinfo{author}{\bibfnamefont{C.~A.} \bibnamefont{Rodriguez-Rosario}},
  \bibnamefont{and}
  \bibinfo{author}{\bibfnamefont{A.}~\bibnamefont{Aspuru-Guzik}},
  \bibinfo{journal}{Phys. Rev. A} \textbf{\bibinfo{volume}{81}},
  \bibinfo{pages}{022323} (\bibinfo{year}{2010}).

\bibitem[{\citenamefont{Sinayskiy et~al.}(2012)\citenamefont{Sinayskiy, Marais,
  Petruccione, and Ekert}}]{SMPE12}
\bibinfo{author}{\bibfnamefont{I.}~\bibnamefont{Sinayskiy}},
  \bibinfo{author}{\bibfnamefont{A.}~\bibnamefont{Marais}},
  \bibinfo{author}{\bibfnamefont{F.}~\bibnamefont{Petruccione}},
  \bibnamefont{and} \bibinfo{author}{\bibfnamefont{A.}~\bibnamefont{Ekert}},
  \bibinfo{journal}{Phys. Rev. Lett.} \textbf{\bibinfo{volume}{108}},
  \bibinfo{pages}{020602} (\bibinfo{year}{2012}).

\bibitem[{\citenamefont{{Mohseni} et~al.}(2008)\citenamefont{{Mohseni},
  {Rebentrost}, {Lloyd}, and {Aspuru-Guzik}}}]{MRLA08}
\bibinfo{author}{\bibfnamefont{M.}~\bibnamefont{{Mohseni}}},
  \bibinfo{author}{\bibfnamefont{P.}~\bibnamefont{{Rebentrost}}},
  \bibinfo{author}{\bibfnamefont{S.}~\bibnamefont{{Lloyd}}}, \bibnamefont{and}
  \bibinfo{author}{\bibfnamefont{A.}~\bibnamefont{{Aspuru-Guzik}}},
  \bibinfo{journal}{J. Chem. Phys.} \textbf{\bibinfo{volume}{129}},
  \bibinfo{pages}{174106} (\bibinfo{year}{2008}).

\bibitem[{\citenamefont{Lloyd et~al.}(2011)\citenamefont{Lloyd, Mohseni,
  Shabani, and Rabitz}}]{lloyd2011}
\bibinfo{author}{\bibfnamefont{S.}~\bibnamefont{Lloyd}},
  \bibinfo{author}{\bibfnamefont{M.}~\bibnamefont{Mohseni}},
  \bibinfo{author}{\bibfnamefont{A.}~\bibnamefont{Shabani}}, \bibnamefont{and}
  \bibinfo{author}{\bibfnamefont{H.}~\bibnamefont{Rabitz}},
  \bibinfo{journal}{arXiv preprint:1111.4982}  (\bibinfo{year}{2011}).

\bibitem[{\citenamefont{Wigner}(1959)}]{W31}
\bibinfo{author}{\bibfnamefont{E.~P.} \bibnamefont{Wigner}},
  \emph{\bibinfo{title}{Group Theory and its Application to the Quantum
  Mechanics of Atomic Spectra}} (\bibinfo{publisher}{New York: Academic Press},
  \bibinfo{year}{1959}), \bibinfo{note}{translation by J. J. Griffin of 1931,
  Gruppentheorie und ihre Anwendungen auf die Quantenmechanik der Atomspektren,
  Vieweg Verlag, Braunschweig.}

\bibitem[{\citenamefont{Datta}(2005)}]{Datta}
\bibinfo{author}{\bibfnamefont{S.}~\bibnamefont{Datta}},
  \emph{\bibinfo{title}{Quantum transport: atom to transistor}}
  (\bibinfo{publisher}{Cambridge University Press}, \bibinfo{year}{2005}).

\bibitem[{\citenamefont{Caruso et~al.}(2009)\citenamefont{Caruso, Chin, Datta,
  Huelga, and Plenio}}]{caruso09}
\bibinfo{author}{\bibfnamefont{F.}~\bibnamefont{Caruso}},
  \bibinfo{author}{\bibfnamefont{A.~W.} \bibnamefont{Chin}},
  \bibinfo{author}{\bibfnamefont{A.}~\bibnamefont{Datta}},
  \bibinfo{author}{\bibfnamefont{S.~F.} \bibnamefont{Huelga}},
  \bibnamefont{and} \bibinfo{author}{\bibfnamefont{M.~B.}
  \bibnamefont{Plenio}}, \bibinfo{journal}{The Journal of Chemical Physics}
  \textbf{\bibinfo{volume}{131}}, \bibinfo{eid}{105106}
  (pages~\bibinfo{numpages}{15}) (\bibinfo{year}{2009}).

\bibitem[{\citenamefont{Sarovar et~al.}(2010)\citenamefont{Sarovar, Ishizaki,
  Fleming, and Whaley}}]{fleming10}
\bibinfo{author}{\bibfnamefont{M.}~\bibnamefont{Sarovar}},
  \bibinfo{author}{\bibfnamefont{A.}~\bibnamefont{Ishizaki}},
  \bibinfo{author}{\bibfnamefont{G.~R.} \bibnamefont{Fleming}},
  \bibnamefont{and} \bibinfo{author}{\bibfnamefont{K.~B.}
  \bibnamefont{Whaley}}, \bibinfo{journal}{Nature Physics}
  \textbf{\bibinfo{volume}{6}}, \bibinfo{pages}{462} (\bibinfo{year}{2010}),
  ISSN \bibinfo{issn}{1745-2473}.

\bibitem[{\citenamefont{Ringsmuth et~al.}(2012)\citenamefont{Ringsmuth,
  Milburn, and Stace}}]{ringsmuth2012}
\bibinfo{author}{\bibfnamefont{A.}~\bibnamefont{Ringsmuth}},
  \bibinfo{author}{\bibfnamefont{G.}~\bibnamefont{Milburn}}, \bibnamefont{and}
  \bibinfo{author}{\bibfnamefont{T.}~\bibnamefont{Stace}},
  \bibinfo{journal}{Nature Physics} \textbf{\bibinfo{volume}{8}},
  \bibinfo{pages}{562} (\bibinfo{year}{2012}).

\bibitem[{\citenamefont{Plenio and Huelga}(2008)}]{plenio08}
\bibinfo{author}{\bibfnamefont{M.~B.} \bibnamefont{Plenio}} \bibnamefont{and}
  \bibinfo{author}{\bibfnamefont{S.~F.} \bibnamefont{Huelga}},
  \bibinfo{journal}{New Journal of Physics} \textbf{\bibinfo{volume}{10}},
  \bibinfo{pages}{113019} (\bibinfo{year}{2008}).

\bibitem[{\citenamefont{Ac{\'\i}n et~al.}(2007)\citenamefont{Ac{\'\i}n, Cirac,
  and Lewenstein}}]{Acin}
\bibinfo{author}{\bibfnamefont{A.}~\bibnamefont{Ac{\'\i}n}},
  \bibinfo{author}{\bibfnamefont{J.~I.} \bibnamefont{Cirac}}, \bibnamefont{and}
  \bibinfo{author}{\bibfnamefont{M.}~\bibnamefont{Lewenstein}},
  \bibinfo{journal}{Nature Physics} \textbf{\bibinfo{volume}{3}},
  \bibinfo{pages}{256} (\bibinfo{year}{2007}).

\bibitem[{\citenamefont{Perseguers et~al.}(2010)\citenamefont{Perseguers,
  Lewenstein, Ac{\'\i}n, and Cirac}}]{Acin2}
\bibinfo{author}{\bibfnamefont{S.}~\bibnamefont{Perseguers}},
  \bibinfo{author}{\bibfnamefont{M.}~\bibnamefont{Lewenstein}},
  \bibinfo{author}{\bibfnamefont{A.}~\bibnamefont{Ac{\'\i}n}},
  \bibnamefont{and} \bibinfo{author}{\bibfnamefont{J.}~\bibnamefont{Cirac}},
  \bibinfo{journal}{Nature Physics} \textbf{\bibinfo{volume}{6}},
  \bibinfo{pages}{539} (\bibinfo{year}{2010}).

\bibitem[{\citenamefont{{Watts} and {Strogatz}}(1998)}]{WS98}
\bibinfo{author}{\bibfnamefont{D.~J.} \bibnamefont{{Watts}}} \bibnamefont{and}
  \bibinfo{author}{\bibfnamefont{S.~H.} \bibnamefont{{Strogatz}}},
  \bibinfo{journal}{Nature} \textbf{\bibinfo{volume}{393}},
  \bibinfo{pages}{409} (\bibinfo{year}{1998}).

\bibitem[{\citenamefont{Erd\H{o}s and R\'{e}nyi}(1960)}]{ER60}
\bibinfo{author}{\bibfnamefont{P.}~\bibnamefont{Erd\H{o}s}} \bibnamefont{and}
  \bibinfo{author}{\bibfnamefont{A.}~\bibnamefont{R\'{e}nyi}}, in
  \emph{\bibinfo{booktitle}{Publication of the Mathematical Institute of the
  Hungarian Academy of Sciences}} (\bibinfo{year}{1960}), pp.
  \bibinfo{pages}{17--61}.

\bibitem[{\citenamefont{Barab\'asi and Albert}(1999)}]{BA99}
\bibinfo{author}{\bibfnamefont{A.-L.} \bibnamefont{Barab\'asi}}
  \bibnamefont{and} \bibinfo{author}{\bibfnamefont{R.}~\bibnamefont{Albert}},
  \bibinfo{journal}{Science} \textbf{\bibinfo{volume}{286}},
  \bibinfo{pages}{509} (\bibinfo{year}{1999}).

\bibitem[{\citenamefont{Bergholm}(2009)}]{Ville}
\bibinfo{author}{\bibfnamefont{V.}~\bibnamefont{Bergholm}},
  \bibinfo{journal}{Quantum Information Toolkit for {MATLAB}}
  (\bibinfo{year}{2009}), \urlprefix\url{http://qit.sourceforge.net/}.

\bibitem[{\citenamefont{Z\"ahringer et~al.}(2010)\citenamefont{Z\"ahringer,
  Kirchmair, Gerritsma, Solano, Blatt, and Roos}}]{PhysRevLett.104.100503}
\bibinfo{author}{\bibfnamefont{F.}~\bibnamefont{Z\"ahringer}},
  \bibinfo{author}{\bibfnamefont{G.}~\bibnamefont{Kirchmair}},
  \bibinfo{author}{\bibfnamefont{R.}~\bibnamefont{Gerritsma}},
  \bibinfo{author}{\bibfnamefont{E.}~\bibnamefont{Solano}},
  \bibinfo{author}{\bibfnamefont{R.}~\bibnamefont{Blatt}}, \bibnamefont{and}
  \bibinfo{author}{\bibfnamefont{C.~F.} \bibnamefont{Roos}},
  \bibinfo{journal}{Phys. Rev. Lett.} \textbf{\bibinfo{volume}{104}},
  \bibinfo{pages}{100503} (\bibinfo{year}{2010}).

\bibitem[{\citenamefont{Lanyon et~al.}(2011)\citenamefont{Lanyon, Hempel, Nigg,
  M{\"u}ller, Gerritsma, Z{\"a}hringer, Schindler, Barreiro, Rambach, Kirchmair
  et~al.}}]{lanyon}
\bibinfo{author}{\bibfnamefont{B.~P.} \bibnamefont{Lanyon}},
  \bibinfo{author}{\bibfnamefont{C.}~\bibnamefont{Hempel}},
  \bibinfo{author}{\bibfnamefont{D.}~\bibnamefont{Nigg}},
  \bibinfo{author}{\bibfnamefont{M.}~\bibnamefont{M{\"u}ller}},
  \bibinfo{author}{\bibfnamefont{R.}~\bibnamefont{Gerritsma}},
  \bibinfo{author}{\bibfnamefont{F.}~\bibnamefont{Z{\"a}hringer}},
  \bibinfo{author}{\bibfnamefont{P.}~\bibnamefont{Schindler}},
  \bibinfo{author}{\bibfnamefont{J.~T.} \bibnamefont{Barreiro}},
  \bibinfo{author}{\bibfnamefont{M.}~\bibnamefont{Rambach}},
  \bibinfo{author}{\bibfnamefont{G.}~\bibnamefont{Kirchmair}},
  \bibnamefont{et~al.}, \bibinfo{journal}{Science}
  \textbf{\bibinfo{volume}{334}}, \bibinfo{pages}{57} (\bibinfo{year}{2011}).

\bibitem[{\citenamefont{Kim et~al.}(2010)\citenamefont{Kim, Chang, Korenblit,
  Islam, Edwards, Freericks, Lin, Duan, and Monroe}}]{monroenature}
\bibinfo{author}{\bibfnamefont{K.}~\bibnamefont{Kim}},
  \bibinfo{author}{\bibfnamefont{M.~S.} \bibnamefont{Chang}},
  \bibinfo{author}{\bibfnamefont{S.}~\bibnamefont{Korenblit}},
  \bibinfo{author}{\bibfnamefont{R.}~\bibnamefont{Islam}},
  \bibinfo{author}{\bibfnamefont{E.~E.} \bibnamefont{Edwards}},
  \bibinfo{author}{\bibfnamefont{J.~K.} \bibnamefont{Freericks}},
  \bibinfo{author}{\bibfnamefont{G.~D.} \bibnamefont{Lin}},
  \bibinfo{author}{\bibfnamefont{L.~M.} \bibnamefont{Duan}}, \bibnamefont{and}
  \bibinfo{author}{\bibfnamefont{C.}~\bibnamefont{Monroe}},
  \bibinfo{journal}{Nature} \textbf{\bibinfo{volume}{465}},
  \bibinfo{pages}{590} (\bibinfo{year}{2010}).

\bibitem[{\citenamefont{Kim et~al.}(2011)\citenamefont{Kim, Korenblit, Islam,
  Edwards, Chang, Noh, Carmichael, Lin, Duan, Wang et~al.}}]{monroetheory}
\bibinfo{author}{\bibfnamefont{K.}~\bibnamefont{Kim}},
  \bibinfo{author}{\bibfnamefont{S.}~\bibnamefont{Korenblit}},
  \bibinfo{author}{\bibfnamefont{R.}~\bibnamefont{Islam}},
  \bibinfo{author}{\bibfnamefont{E.~E.} \bibnamefont{Edwards}},
  \bibinfo{author}{\bibfnamefont{M.-S.} \bibnamefont{Chang}},
  \bibinfo{author}{\bibfnamefont{C.}~\bibnamefont{Noh}},
  \bibinfo{author}{\bibfnamefont{H.}~\bibnamefont{Carmichael}},
  \bibinfo{author}{\bibfnamefont{G.-D.} \bibnamefont{Lin}},
  \bibinfo{author}{\bibfnamefont{L.-M.} \bibnamefont{Duan}},
  \bibinfo{author}{\bibfnamefont{C.~C.~J.} \bibnamefont{Wang}},
  \bibnamefont{et~al.}, \bibinfo{journal}{New J. Phys.}
  \textbf{\bibinfo{volume}{13}}, \bibinfo{pages}{105003}
  (\bibinfo{year}{2011}).

\bibitem[{\citenamefont{Kirchmair et~al.}(2009)\citenamefont{Kirchmair,
  Benhelm, Z{\"a}hringer, Gerritsma, Roos, and Blatt}}]{thermalkirchmair}
\bibinfo{author}{\bibfnamefont{G.}~\bibnamefont{Kirchmair}},
  \bibinfo{author}{\bibfnamefont{J.}~\bibnamefont{Benhelm}},
  \bibinfo{author}{\bibfnamefont{F.}~\bibnamefont{Z{\"a}hringer}},
  \bibinfo{author}{\bibfnamefont{R.}~\bibnamefont{Gerritsma}},
  \bibinfo{author}{\bibfnamefont{C.~F.} \bibnamefont{Roos}}, \bibnamefont{and}
  \bibinfo{author}{\bibfnamefont{R.}~\bibnamefont{Blatt}},
  \bibinfo{journal}{New Journal of Physics} \textbf{\bibinfo{volume}{11}},
  \bibinfo{pages}{023002} (\bibinfo{year}{2009}).

\bibitem[{\citenamefont{Adolphs and Renger}(2006)}]{renger2006}
\bibinfo{author}{\bibfnamefont{J.}~\bibnamefont{Adolphs}} \bibnamefont{and}
  \bibinfo{author}{\bibfnamefont{T.}~\bibnamefont{Renger}},
  \bibinfo{journal}{Biophys. J.} \textbf{\bibinfo{volume}{91}},
  \bibinfo{pages}{2778} (\bibinfo{year}{2006}).

\end{thebibliography}

\section*{Additional Information}
\begin{description}
  \item [Acknowledgements]\noindent
    We thank Michele Allegra, Stephen Clark, Seth Lloyd and Ville Bergholm for helpful comments on the draft.
    Parts of this work were supported by the European Commission under grants COQUIT, ERC Grant GEDENTQOPT, and CHISTERA QUASAR.
\end{description}

\appendix

\section*{Supplementary Information}

\subsection*{ S1 Analytic examples for enhancement, suppression and direction} 

\begin{figure*}
  \begin{center}
        {\includegraphics[width=0.8\textwidth]{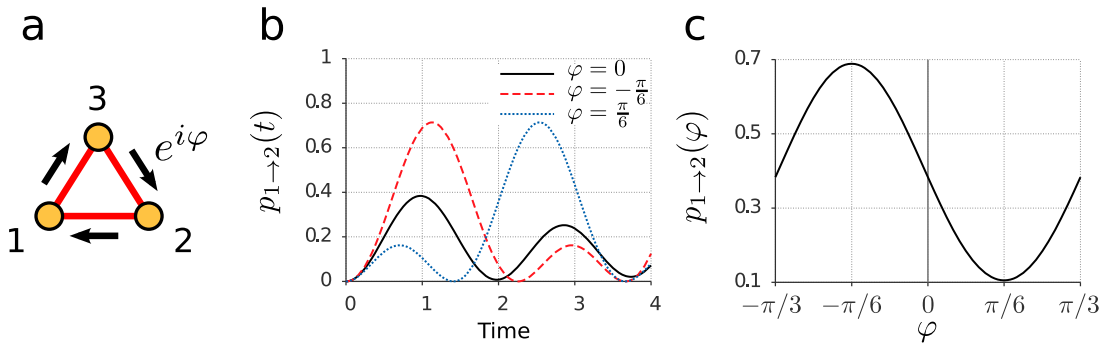}}
  \end{center}
  \caption{
  ({\bf A}) Triangle with inhomogeneous coupling and chiral phases.
  ({\bf B}) Illustrates the quantity $P_{1\to 2}(t)$  for the triangle graph with
  inhomogeneous coupling $J_{12}=1, J_{23}=1.3, J_{13}=0.5$. 
  Directing the phases of the chiral walker effects the transport, in
  particular, there is enhancement for $\theta= 3\varphi=\pi/2$ in both 
  arrival time and maximal probability. 
  ({\bf C}) The
  probability varies as a function of the chiral angle at a given time near the
  first maximum.}
  \label{fig:triangle}
\end{figure*}

Using the example of the regular polygon (that has $J_{mn}\equiv 1$ for all edges) 
we will explain how enhancement, suppression and direction of transport
can be influenced by the phases in chiral quantum walks.
Typically we need to solve the eigenvalue problem of the Hamiltonian 
$H \eket{j}=E_j\eket{j}$ (the tilde denotes energy eigenstates) to compute 
\begin{equation*}
  P_{S\to E}(t)=
  \left|\sum_{\tilde{j}}^N \langle E|\tilde{j}\rangle\langle\tilde{j}|S\rangle e^{-iE_j t}\right|^2
\end{equation*}
For a regular homogeneous polygon 
($H=\sum_{n=1}^N e^{i\varphi} \ket{n}\bra{n+1}+h.c.$ with cyclic boundary condition) 
the eigenvectors and
eigenvalues are 
found from the Fourier Transform
\begin{equation*}
  \eket{k}=\frac{1}{\sqrt{N}}\sum_{n=1}^{N}
  e^{-\frac{2\pi\,i\,k\,n}{N}}\ket{n}\ ,\quad
  E_k(\varphi)=2\cos\left(\frac{2\pi k}{N}-\varphi\right)\ ,
\end{equation*}
The \stsp\ from site $S$ to $E$ then reads
\begin{equation}
  P_{S\rightarrow E}(t, \varphi) =
\left| \frac 1N \sum_{k=0}^{N-1} e^{-i\left(2t\cos(\frac{2\pi k}{N}-\varphi)+\frac{2\pi k(E-S)}{N}\right)}\right|^2\label{regpoly}
\end{equation}
We now list several features
\begin{itemize}
\item Direction and enhancement

In the case of time symmetric quantum walks,
for the homogeneous polygon $P_{S\to E}=P_{S\to N-E+2S}$.
This implies that if the number of edges is odd, the \stsp\ 
is at most $1/2$, since there is always a vertex different from $E$ 
at which the probability is identical. 
The chiral case breaks this symmetry.
In particular, for the regular triangle one can reach
$P_{1\to 2}(t)=1$; one can find values of $\theta$ analytically by
equating the phases of the three exponentials in the \stsp\ modulo $2\pi$ and solve the 
two linear equations for $t$ and $\theta$. Enhancement and direction is 
indicated in Fig.~\ref{fig:triangle}, where a triangle graph is shown with 
inhomogeneous coupling. 
\item Complete suppression of transport

Complete suppression of transport is possible in loops with an even number of sites.  They belong
to the set of {\em bipartite} graphs, whose vertices can be
partitioned into two sets such that edges connect only members from
different subsets. Directing transport in bipartite graphs is
forbidden. This can be easily shown by noting that the gauge transformation
$\ket{n}\mapsto -\ket{n}$ for all members of one subset
has the effect of $H\mapsto-H$, which is precisely the time reversal transformation. 
Hence 
\begin{align*}
  P_{S\to E}(t)&=
  P_{S\to E}(-t)=
  \mbox{Tr}(e^{iHt}\rho_S e^{-iHt}\rho_E)\\
  &=
  \mbox{Tr}(e^{-iHt}\rho_E e^{iHt}\rho_S)=
  P_{E\to S}(t)\ .
\end{align*}
Even
chiral loops suppress transport as
$\theta=\pi$ results in $p_{1\to N/2+1}(t)=0$ for all times $t$.
This can again be shown by writing down the formula \eqref{regpoly} and
separating it into two sums, one for $k$ being even and the other for $k$ odd. 
The simplest even polygon is the quadrilateral that
can be used as a building block for the realization of transport suppressing
topologies.
\end{itemize}

\begin{figure}
  \begin{center}
    \includegraphics[width=0.8\columnwidth]{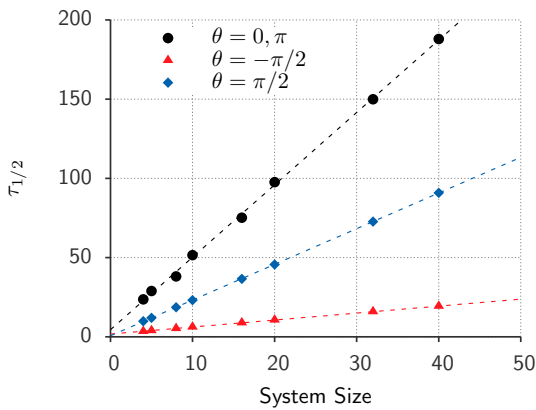}
  \end{center}
  \caption{Half arrival time scaling.
  Dependence of the half arrival time,
  $\tau_{1/2}$, on the length of the triangle chain shown in Figure~2A 
for different values of the phases.
  Both chiral ($\theta\neq 0$) and achiral ($\theta=0$) cases exhibit
 linear scaling behaviour. The slope can be tuned by changing the value of
the phase parameter.
  }
  \label{fig:saw-t12}
\end{figure}

\noindent {\bf Robustness of chiral transport enhancement of the triangle chain.} In the triangle chain discussed in the main text, the half arrival time
$\tau_{1/2}$ depends linearly on the size of the system, see 
Figure~\ref{fig:saw-t12}.
This fact is a characteristic of quantum walks in linear chains and
continues to hold for the chiral triangle chain.

\subsection*{S2 Experimental proposal}

In this section we outline a proposal to simulate chiral quantum walks in a system of ultra-cold trapped atomic ions. 
The proposal can be realized with currently available techniques and technology. 
Trapped ion systems have previously been used to investigate non-chiral quantum walks \cite{PhysRevLett.104.100503}, where the motional state was used to encode the walk. 
In contrast, in our proposal both chiral and non-chiral walks are encoded into the electronic state of the ions. 
The walk dynamics is generated by ion-ion interactions mediated by joint vibrational modes of the ion string. 
These interactions can be driven by laser-induced optical dipole forces, for example, and can be implemented with high quality as shown in several recent works \cite{lanyon, monroenature}.  In the text below we first introduce the 3-ion system that the experiment requires. Then we explain how to engineer the quantum walk Hamiltonians that we wish to investigate. Next, we calculate the chiral and non-chiral phenomena that can be generated by these Hamiltonians. Finally, we outline the experimental procedure and note some important experimental considerations. 

\noindent{\bf Implementation in trapped ions.} We consider a string of three ions in a linear ion crystal, which can be achieved with standard linear-Paul traps. A long-lived electronic transition
inside each ion encodes a two-level spin. The mapping could be that spin-down
is represented by an ion in the ground state, while spin-up is represented by
an ion in the excited state of this transition, for example. The 8 possible
logical states of this system then become
$\ket{\downarrow,\downarrow,\downarrow},\ket{\downarrow,\downarrow,\uparrow}, \ket{\downarrow,\uparrow,\downarrow},......,\ket{\uparrow,\uparrow,\uparrow}$. 
In our proposal a subset of these states represents the sites of the quantum
walk, as will be described. The dynamics of any quantum walk between these states require the ability to turn on interactions between the ionic spins. Effective ion-ion interactions can be achieved by off-resonantly coupling the electronic states in which the spin is encoded to one or more common vibrational modes of the string. Such a spin-dependant force can be implemented using a laser field with two symmetrically tuned frequencies at $\omega_s\pm\mu$ with $\mu\ll\omega_s$. If the symmetric detuning is sufficiently far from all motional side-bands so that the generation of phonons can be adiabatically eliminated \cite{monroetheory}, then a pure spin-spin interaction is generated of the form:
\begin{equation}
H(\phi)=\sum_{i<j}J_{ij}\sigma_{\phi}^{i}\sigma_{\phi}^{j}\ ,
\end{equation}
where 
\begin{equation*}
  \sigma_{\phi}^i{=}\cos{\phi}\,\sigma_x^i+\sin{\phi}\sigma_y^i\label{rots}
\end{equation*}
\begin{equation}
J_{ij}=\frac{\Omega^2\eta^2\omega}{\Delta^2-\omega^2}b_ib_j
\end{equation}
Here $\Omega$ describes the laser coupling strength to the electronic
transition, $\eta_i$ is the Lamb-Dicke parameter describing the coupling
strength between the laser and the motional side-band for a single ion. An experimentally tunable phase $\phi$ is set by the phase difference of the two driving laser fields. Finally, $b_{i}$ describes how strongly ion $i$ couples to motional mode $\omega$. 
The increasing difficulty of generating spin-spin interactions with more ions in a string is incorporated in $b_{i}$. 

In this proposal we will simultaneously use two of the normal modes along the axis of the string, called the centre-of-mass (COM) and breathing (Br) modes, to mediate spin-spin interactions. These modes are typically well-spaced in frequency and can therefore can be individually addressed. The coupling vectors $b_i$ for the axial COM and breathing are $1/\sqrt{3}\times[1,1,1]$ and $1/\sqrt{2}\times[1,0,-1]$, respectively. Consequently, the contributions to the spin-spin interaction Hamiltonian of these modes are:
\begin{eqnarray}
H_{COM}(\phi_1)&=&J_{COM}~(\sigma_{\phi_1}^{1}\sigma_{\phi_1}^{2}
+\sigma_{\phi_1}^{2}\sigma_{\phi_1}^{3}
+\sigma_{\phi_1}^{1}\sigma_{\phi_1}^{3}) \label{egy}\\
H_{Br}(\phi_2)&=&J_{Br}~\sigma_{\phi_2}^{1}\sigma_{\phi_2}^{3}\label{ketto}
\end{eqnarray}
where 
\begin{equation}
J_i=\frac{\Omega_i^2\eta_i^2\omega_i}{\Delta_i^2-\omega_i^2}\label{eqJ}
\end{equation}
determines the coupling strength of the spin-spin interaction due to vibrational mode $i$.  This equation allows for separate pairs of light fields to drive each mode  simultaneously, which can therefore have different coupling strengths and detunings.  This is required because we wish to be able to precisely control the relative strengths $A_i$, which can be achieve by changing the relative strengths of the laser fields. Separate pairs of light fields are also necessary since we require the ability to individually set the phase $\phi_i$ determined by the phase difference between the laser fields in each pair. 

\begin{figure}
  \begin{center}
    \includegraphics[width=\columnwidth]{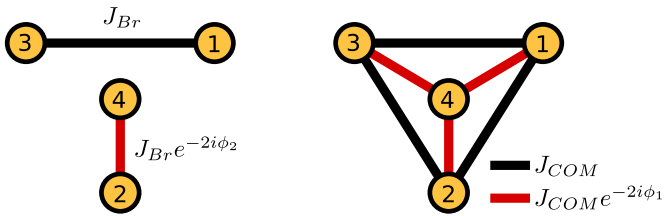}
  \end{center}
  \caption{Networks describing the modes $H_{Br}$ (left) and $H_{COM}$ (right).
  }
  \label{fig:exp-couplings}
\end{figure}

\noindent{\bf Ideal experimental behavior of the proposed walks.}
Now that we have established the Hamiltonians to drive the quantum walks, given
by equations \eqref{egy}  and \eqref{ketto}, we will examine the walk 
dynamics that we wish to explore. 
A key observation is that the operators $J\,\sigma^m_\phi \sigma^n_\phi\;(m,n\in\{1,2,3\}, m\neq n$)
leaves invariant the subspace spanned by the following four spin states 
\begin{align*}
\ket{1}\equiv\ket{\uparrow,\downarrow,\downarrow},&
\quad\ket{2}\equiv\ket{\downarrow,\uparrow,\downarrow},\quad\\ 
\ket{3}\equiv\ket{\downarrow,\downarrow,\uparrow},&
\quad \ket{4}\equiv\ket{\uparrow,\uparrow,\uparrow}\ .
\end{align*}
In order to see this, 
consider the following decomposition ($m, n\in\{1,2,3\},\;m\neq n$):
\begin{equation}
\sigma^m_\phi \sigma^n_\phi=
\left(\cos\phi\sigma^m_x +\sin\phi\sigma^m_y \right)
\left(\cos\phi\sigma^n_x +\sin\phi\sigma^n_y \right)
\end{equation}
in terms of the creation and annihilation operators
$\sigma^m_\pm=\sigma^m_x \pm i\sigma^m_y$.
It follows that:
\begin{equation}
  \sigma^m_\phi \sigma^n_\phi=
    e^{-2i\phi}\sigma^m_+\sigma^n_+ +
    e^{2i\phi}\sigma^m_-\sigma^n_- +
    \sigma^m_+\sigma^n_- +\sigma^m_-\sigma^n_+
\end{equation}
Here
$\ket{n}$ and $\ket{m}$ are coupled by the real strength $J$, whereas $\ket{4}$ is coupled
to the remaining site by the same strength multiplied with the phase
$e^{-2i\phi}$. Now, notice that the modes $H_{COM}$ and $H_{Br}$ are 
built from operators of the above form.
Using these models we can realize quantum
walks on the four sites given above. 
Now one determines the parameters of the quantum walk Hamiltonians:
\begin{equation*}
H^{(k)}=  \sum_k\sum_{n,m=1}^{4} J^{(k)}_{nm}|n\rangle\langle m|\quad\mbox{with}\quad J^{(k)}_{nm}=\overline{J}^{(k)}_{mn}\ ,
 \end{equation*}
where the index $k$ refers to the two different modes  $H_{COM}(\phi_1)$ and
$H_{Br}(\phi_2)$. Figure~\ref{fig:exp-couplings} shows the 
general situation.

The values to be used for the experiment are
given in the following table:
\begin{center}
  \begin{tabular}{lrrrr}
    \toprule
     & $J_{COM}$ & $J_{Br}$ & $\phi_1$ & $\phi_2$\\
    \colrule
    \em $H_{CQW_1}$&$2$&$-3$&$\pi/2$&$0.304\,\pi$\\
    \em $H_{CQW_2}$&$2$&$-3$&$\pi/2$&$-0.304\,\pi$\\
    \em $H_{QW}$&$-2$&$1$&$\pi/2$&$0$ \\
    \botrule
  \end{tabular}
\end{center}

\begin{figure}
\begin{center}
\includegraphics[width=\columnwidth]{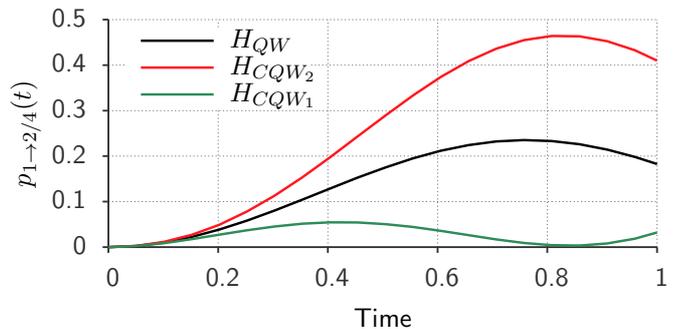}
\end{center}
\caption{The effect of the chiral phase parameter on the occupation probability.
The plot shows the chiral walkers at site 2 and the achiral one at site 4, all
being initially at site 1. By this arrangement the real coupling strengths of the
corresponding Hamiltonians
are the same.}\label{plotsforexp}
\end{figure}

The parameter $\phi_2$ is determined by $\cos 2\phi_2=-1/3$.
Since the effect $\phi_1$ is just a minus sign in $H_{COM}$ for the edges
incident to $\ket{4}$,  we see that $H_{CQW_1}$ and $H_{CQW_2}$ are the 
images of each other under TRS transformation. One can also check that 
$H_{QW}$ is the time-reversal symmetric counterpart of the chiral walks  in that all its 
couplings are real and equal to the absolute values of the chiral walks after we relabel sites 
$\ket{2}$  and $\ket{4}$. Computer simulation (Figure~\ref{plotsforexp}) shows the effect of TRS breaking, since $P_{1\to 2}(t)\neq P_{2\to 1}(t)$, 
enhancement and suppression of transport as the first maxima of $H_{CQW_2}$ and
$H_{CQW_1}$ are of considerably larger (smaller)
magnitude, respectively, than that of the corresponding symmetric walker. When we compute $P_{2\to 1}(t)$, the roles of the two chiral walks are
exchanged due to the symmetry of the model.  

\vspace{2mm}
\noindent{\bf Experimental details.} As a specific example we consider using three $^{40}$Ca$^+$ ions in a standard linear Paul trap. This atomic species is used by several groups around the world and can be precisely manipulated in the way that we require. Spins can be encoded into a metastable electric quadrupole transition, the excited state of which has a lifetime of approximately 1 second which is much longer than the times required for quantum operations. Spin operations can be driven using a 729nm laser, spin-spin interactions can be implemented using a bichromatic light field with symmetrical detuned side-bands around the electronic transition \cite{thermalkirchmair}. We note that implementation of our proposal using another atomic species with a hyperfine transition, for example, would also be possible. 

\begin{figure}
\begin{center} 
\includegraphics[width=0.9\columnwidth]{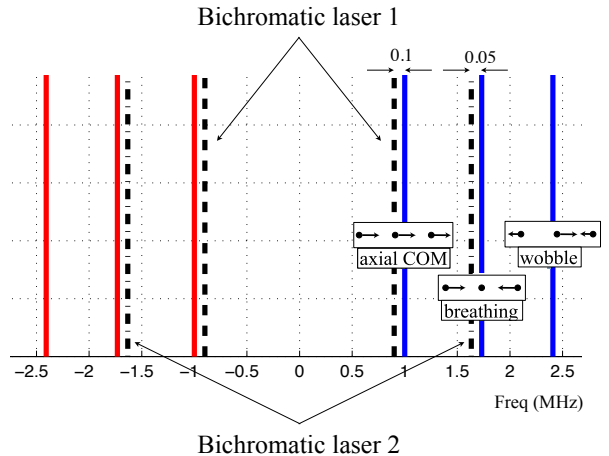}
\end{center} 
\caption{Mode frequency spectrum of a linear three ion crystal of $^{40}$Ca$^+$ in a typical harmonic trap with radial COM (not shown) and axial COM frequencies of 5 MHz and 1MHz, respectively. Two bichromatic laser fields, each symmetrically detuned from the axial COM (by 100KHz) and breathing modes (by 50KHz), drive interactions between spins encoded into an electronic transition of each ion. The phase difference between the spectral components in bichromatic field 1 (2) determines the phase $\phi$ in the Hamiltonian given in equation \eqref{egy}  (\eqref{ketto}).
}\label{longrange}
\end{figure}

Figure~\ref{longrange} shows one possibility allowing simultaneous realisation of the Hamiltonians given in equations \eqref{egy} and \eqref{ketto}. Two pairs of bichromatic fields are turned on, on pair tuned close to the axial COM and another to the axial breathing mode. We choose to use the axial modes, since they are well-spaced out in frequency, thereby enabling each bichromatic light field to be simultaneously close enough to its nearest vibrational mode such that the effect of that mode completely dominates the spin-spin interaction and sufficiently far away such that the adiabatic approximation still applies.

The criterion to generate a pure spin-spin interaction with each bichromatic field is $|\omega_m-\mu|\gg\eta_m\omega_m$, i.e. the detuning from all side-bands is much greater than the coupling strength on those modes. Using the detunings shown in Figure~\ref{longrange}, and $\Omega{=}2\pi\times 100KHz$ we obtain ratios $|\omega_m-\mu|/\eta_m\omega_m$ of 21 and 14 for the COM and breathing bichromats, respectively. In this regime, errors due to the generation of phonons are extremely small. 
Regarding the issue of off-resonant coupling to unwanted modes: straightforward calculations using equation \eqref{eqJ} and the frequencies shown in figure~\ref{longrange} show that in both cases the size of the far-offresonant spin-spin coupling, due to bichromatic field 2 on the COM for example, is more than one order of magnitude less than the desired couplings in every case. This could be further reduced by increasing the frequency separation between axial modes by increasing the confining potential in this direction, for example. 

An important experimental consideration is the requirement to maintain a fixed phase relationship between the two pairs of light fields generating the walk Hamiltonians. For example, this can be achieved by generating them in the same acousto-optic modulator (AOM) and thereby keeping their paths common mode, between the point of generation and the point of interaction with the ions. The maximum frequency splitting is approximately 3.4 MHz, which will result in some angular divergence of the different frequencies at the output facet of the AOM crystal, which can be compensated with linear optics allowing coupling into an optical fibre for transport to the ion trap itself. 

The experiment would proceed as follows. Firstly, standard methods of doppler cooling, resolved side-band cooling on the axial COM and breathing modes, and optical pumping  prepares the three ion string into an ultra low entropy state and the initial spin state $\ket{\downarrow,\downarrow,\downarrow}$. Next, a standard combination of single-ion focused and three-ion focused beams is used to prepare the initial state $\ket{\uparrow,\downarrow,\downarrow}$. The bichromatic light fields which implement the desired walk dynamics are then turned on for a fixed period of time. Finally the state of each encoded spin (up or down) is measured individually using standard fluorescent detection techniques and a CCD camera. Experiments are repeated multiple times from which estimates of the probability for finding spins in all configurations can be deduced.

\subsection*{S3 Fenna-Matthews-Olson complex} 

The Fenna-Matthews-Olson (FMO) complex is a well studied light
harvesting system in green sulphur bacteria.
The interest in FMO resides in the high efficiency shown in the transport of energy from
the antenna to the reaction centre.
The faithful description and simulation of the FMO complex
has been the highlight of recent research. 

The system is composed of three subunits, each made of 
seven BChl-$a$ molecules embedded in a protein scaffold.
We restrict our study to the simulation of the exciton transport in a single
unit.
To simulate the evolution of the system in the one-excitation manifold, we use the
following Liouville equation:

\begin{equation}
\dot\rho=
\mathcal{L}[\rho]=
-i [H,\rho]
+\phi \hat L_\phi\{\rho\}
+\gamma  \hat L_\gamma\{\rho\}
+\tau \hat L_\tau\{\rho\} 
\label{eq:liouville}
\end{equation}
where the Lindblad super-operators are defined as:
\begin{align}
  \hat L_\phi\{\rho\}&\equiv
  \sum_n L_{\phi,n} \rho L_{\phi,n}^\dag
  -\frac 12 \left(L_{\phi,n}^\dag L_{\phi,n}\rho+
          \rho L_{\phi,n}^\dag L_{\phi,n}\right)\\
          L_{\phi,n}&= \ket{n}\bra{n}\label{eq:lind1}\\
  \hat L_\gamma\{\rho\}&\equiv
  \sum_n L_{\gamma,n} \rho L_{\gamma,n}^\dag
  -\frac 12 \left(L_{\gamma,n}^\dag L_{\gamma,n}\rho+
          \rho L_{\gamma,n}^\dag L_{\gamma,n}\right)\\
          L_{\gamma,n}&= \ket{d}\bra{n}\label{eq:lind2}\\
  \hat L_\tau\{\rho\}&\equiv
  \sum_{n\in T} L_{\tau,n} \rho L_{\tau,n}^\dag
  -\frac 12 \left(L_{\tau,n}^\dag L_{\tau,n}\rho+
          \rho L_{\tau,n}^\dag L_{\tau,n}\right)\\
          L_{\tau,n}&=\ket{\tau}\bra{n}\label{eq:lind3}
\end{align}
where the sums are over the site basis and $T$ is the set of sites connected
to the reaction centre (in this case only site three).
The coherent part of the evolution is described using a time reversal symmetric Hamiltonian from the
literature~\cite{renger2006}.
In~\eqref{eq:liouville} the non-coherent terms describe the coupling of the
system with a thermal bath, dephasing, Eq.~\eqref{eq:lind1}, and
recombination, Eq.~\eqref{eq:lind2}, and
the effect of the excitation trapping at the site connected to the reaction
centre, Eq.~\eqref{eq:lind3}.
The thermal bath, for the present paper, is considered as a set of harmonic
oscillators coupled to the system of interest as
in~\cite{MRLA08,Breuer02}.

The initial state is set on site one and the reaction centre is connected as
an energy sink to site three.
In the simulation we use a dephasing rate of $\phi=9.0$ ps$^{-1}$
(which corresponds to 295K),
the recombination rate is $\gamma=1$ ns$^{-1}$ and trapping rate from site three to
the reaction centre is $\tau=1.0$ ps$^{-1}$.

The optimization of the phases was found to be robust with respect to phase changes.
The resulting phases are reported in Table~\ref{tab:phases}.

\begin{table}
  \centering
  \includegraphics[width=0.8\columnwidth]{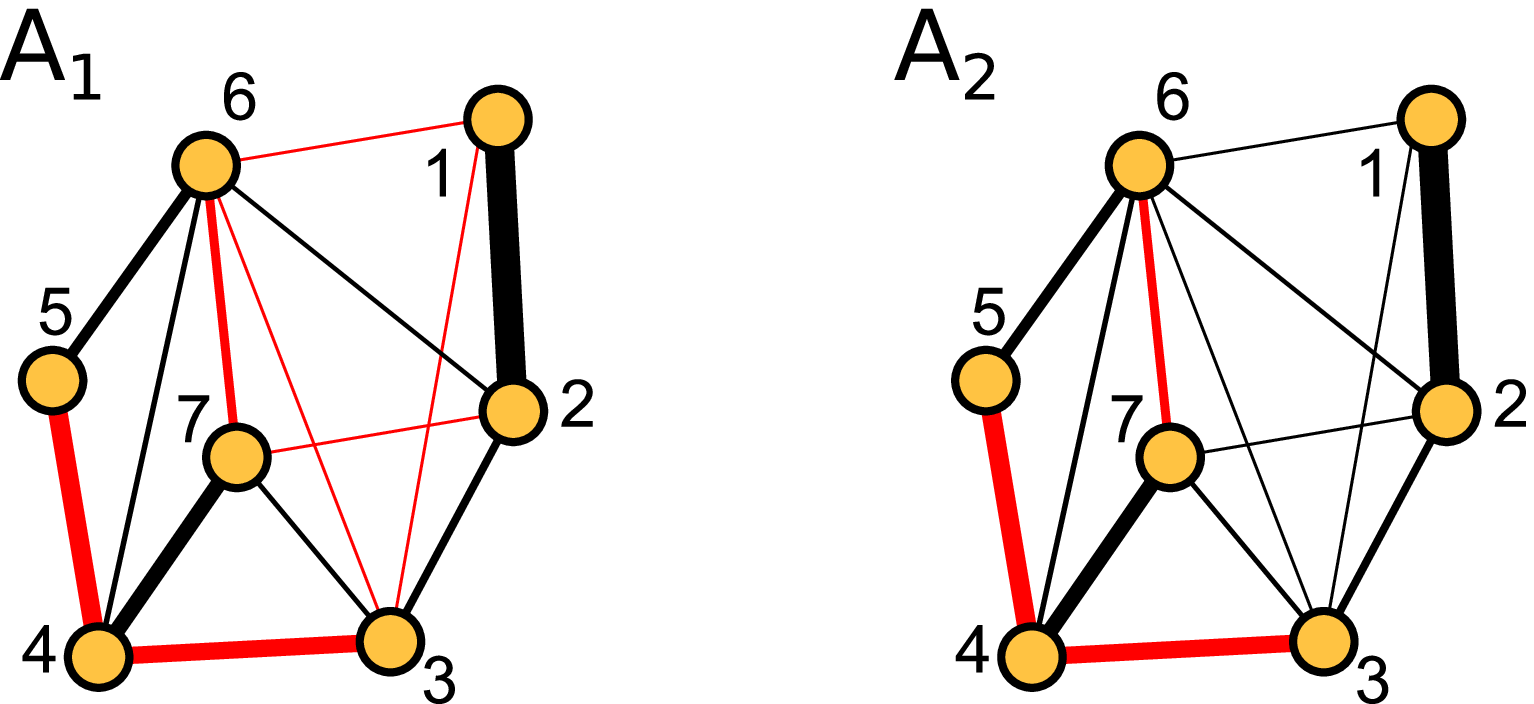}
  \vskip 10pt
  \begin{tabular}{ccccc}
    \toprule
    Edges & $A_1$ & variance & $A_2$ & variance\\
    \colrule
    3-4 & $1.31484899~\pi$ & $4.33\cdot 10^6\pi$ & $1.58371001~\pi$  & $4.86\cdot 10^6\pi$\\
    4-5 & $1.66997830~\pi$ & $5.20\cdot 10^6\pi$ & $1.39551582~\pi$  & $6.60\cdot 10^6\pi$\\
    6-7 & $1.8406103 ~\pi$ & $1.25\cdot 10^5\pi$ & $0.1338368 ~\pi$  & $1.72\cdot 10^5\pi$\\
    2-7 & $1.2949616 ~\pi$ & $2.80\cdot 10^5\pi$\\
    1-6 & $1.67543320~\pi$ & $5.36\cdot 10^6\pi$\\
    1-3 & $0.04222214~\pi$ & $4.36\cdot 10^6\pi$\\
    3-6 & $0.8761298 ~\pi$ & $1.03\cdot 10^5\pi$\\
    \botrule
  \end{tabular}
  \caption{Transport enhancement in the FMO complex.
  Both for $A_1$ and $A_2$, complex phases are applied to the red edges. The results of
  a simultaneous optimization procedure are listed in the table.
  }
  \label{tab:phases}
\end{table}

\subsection*{S4 Small-world networks} 

The Watts-Strogatz model~\cite{WS98} gives a constructive algorithm for 
building a network with 
small-world characteristics, starting from a regular lattice.
The latter is defined as a periodic chain of $N$ nodes where each node is
connected to $k$ neighbours ($k/2$ to the left and $k/2$ to the right).
The final small-world network is obtained by taking all neighbouring edges at each
node and rewiring all the edges toward the node left with a probability $p$.
The limit for $p\rightarrow 1$ of the Watts-Strogatz model is the Erd\H{o}s-R\'enyi random model
with fixed number of edges (restricted to connected graphs).

In our case, we set $N=32$ and $k=4$ and choose only connected
graphs as we are concerned in comparing transport on networks of the same sites.
The initial state is on site $S$ while the sink is a external site connected to site $E$. Site $E$
latter is placed on the opposite side of the initial circle.
The absorption rate of the sink is $r=1.0$.
The evolution is described by the coherent Hamiltonian, which, in this case, 
corresponds to the connectivity matrix of the network.
We define $\tau^{QW}_{1/2}$ the time needed by the normal quantum walker to reach the
probability $p_{sink}(t)=1/2$.
We add phases to the edges neighbouring node $E$ and optimize them in order to
improve the probability of being trapped in the target site at time
$\tau^{QW}_{1/2}$.
The non-coherent part (trapping) is described by a Lindbladian super-operator.
We repeat the evolution for 200 different realizations of the graph for each
value of $p$.

\end{document}